\documentclass[a4paper, 11pt]{amsart}
\usepackage{amsmath, amssymb, amsthm}	
\usepackage{braket, stmaryrd, MnSymbol,ascmac,mathrsfs}
\usepackage{cite}
\usepackage{graphicx}

\calclayout

\newcommand{\pr}[1]{#1^{\prime}}

\newcommand{\del}{\partial}

\newcommand{\mfrak}[1]{\mathfrak{#1}}
\newcommand{\mcal}[1]{\mathcal{#1}}
\newcommand{\mbb}[1]{\mathbb{#1}}
\newcommand{\mrm}[1]{\mathrm{#1}}

\newcommand{\scr}[1]{\mathscr{#1}}
\newcommand{\what}[1]{\widehat{#1}}
\newcommand{\no}[1]{:\hspace{-3pt} #1\hspace{-3pt}:\hspace{3pt}}

\theoremstyle{plain}

\theoremstyle{definition}

\theoremstyle{remark}

\makeatletter
    
    \@addtoreset{equation}{section}
\makeatother

\title{Schramm-Loewner evolution with Lie superalgebra symmetry}
\author{Shinji Koshida}
\address{Department of Basic Science, The University of Tokyo}
\email{koshida@vortex.c.u-tokyo.ac.jp}

\begin{document}

\begin{abstract}
We propose a generalization of Schramm-Loewner evolution (SLE) that has internal degrees of freedom
described by an affine Lie superalgebra.
We give a general formulation of SLE corresponding to representation theory of an affine Lie superalgebra
whose underlying finite dimensional Lie superalgebra is basic classical type,
and write down stochastic differential equations on internal degrees of freedom in case that
the corresponding affine Lie superalgebra is $\what{\mfrak{osp}(1|2)}$.
We also demonstrate computation of local martingales associated with the solution from a representation
of $\what{\mfrak{osp}(1|2)}$.
\end{abstract}

\maketitle

\section{Introduction}
Schramm-Loewner evolution (SLE) \cite{Schramm2000} is
a stochastic process $(g_{t})_{t\ge 0}$ that takes its value in a space of formal power series $z+\mbb{C}[[z^{-1}]]$
and satisfies the following stochastic differential equation (SDE)
\begin{equation}
	\frac{d}{dt}g_{t}(z)=\frac{2}{g_{t}(z)-B_{t}},\ \ \ g_{0}(z)=z,
\end{equation}
where $B_{t}$ is the Brownian motion of covariance $\kappa$ that starts from the origin.
The covariance $\kappa$ of the Brownian motion parametrizes SLEs,
and we often denote the SLE specified by $\kappa$ by SLE$(\kappa)$.
While we regard the solution $g_{t}(z)$ as a formal power series, it actually becomes a uniformizing map of a simply connected domain in the upper half plane.
More precisely, at each time $t$, $g_{t}$ is almost surely a biholomorphic map $g_{t}:\mbb{H}\backslash K_{t}\to \mbb{H}$
for a subset $K_{t}\subset \mbb{H}$, called a hull, in the upper half plane.
Moreover the hulls are increasing, {\it i.e.}, $t<s$ implies $K_{t}\subset K_{s}$.
This means that SLE describes a stochastic evolution of hulls in the upper half plane.
If we look at this evolution of hulls more closely, we find that it is governed by an evolution of the tip of a slit in the upper half plane.
Thus SLE induces a probability measure on a space of curves in the upper half plane running from the origin to the infinity,
which is called the SLE$(\kappa)$-measure.
The SLE$(\kappa)$-measure is proved to describe an interface of clusters in several critical systems in two dimensions \cite{Smirnov2001,ChelkakDuminil-CopinHonglerKemppainenSmirnov2014},
which is why SLE is interesting to the field of statistical physics.
There are many literatures\cite{Lawler2004,RohdeSchramm2005,Werner2003} on SLE from probability theoretical point of view.

Two dimensional critical systems are also studied by means of two dimensional conformal field theory (CFT) \cite{BelavinPolyakovZamolodchikov1984},
which is distinguished from other quantum field theories due to its symmetry described by infinite dimensional Lie algebras.
A milestone was Cardy's formula \cite{Cardy1992}, later proved by Smirnov \cite{Smirnov2001}, which predicts crossing probabilities in the two-dimensional
critical percolation from computation of partition functions for a boundary CFT.

Having two different approaches, SLE and CFT, to two-dimensional critical systems,
one expects that these two notions are related one another in some sense.
The connection between SLE and CFT has been studied under the name of SLE/CFT correspondence \cite{BauerBernard2006,FriedrichWerner2003, FriedrichKalkkinen2004,Friedrich2004,Kontsevich2003,Dubedat2015a,Dubedat2015b,Kytola2007}.
One of significant achievements we follow in this paper is the group theoretical formulation of SLE by Bauer and Bernard \cite{BauerBernard2002, BauerBernard2003a,BauerBernard2003b}.
It regards SLE as a random process that is induced from one on an infinite dimensional Lie group that
governs formal coordinate transformations at infinity,
while the same infinite dimensional Lie group acts on a representation of the Virasoro algebra.
This point of view enables one to compute local martingales associated with SLE from a representation of the Virasoro algebra.

There are several directions of generalization of SLE along SLE/CFT correspondence,
including multiple SLE \cite{BauerBernardKytola2005}, SLE for logarithmic CFT \cite{Rasmussen2004b,Moghimi-AraghiRajabpourRouhani2004} and for the $\mcal{N}=1$ superconformal algebra \cite{Rasmussen2004a}.
CFT with internal degrees of freedom is called Wess-Zumino-Witten (WZW) theory \cite{WessZumino1971, Witten1984, KnizhnikZamolodchikov1984} and it is associated with
the representation theory of an affine Lie algebra.
Variants of SLE corresponding to WZW theory are also considered in literatures \cite{BettelheimGruzbergLudwigWiegmann2005,AlekseevBytskoIzyurov2011,Rasmussen2007,Nazarov2012,SK2017,SK2018a,Sakai2013}.

WZW theory can be generalized to super WZW theory whose internal degrees of freedom are governed by a Lie supergroup,
and there are physics models \cite{Efetov1983,SaleurSchomerus2007, Bernard1995, EsslerFrahmSaleur2005} that are expected to be described by super WZW theory.
It is, however, not known whether there is a generalization of SLE that corresponds to super WZW theory,
and this is the issue we address in this paper.

The achievement of this paper is 
derivation of SDEs whose solution can be regarded as a generalization of SLE
corresponding to super WZW theory.
Let us sketch the strategy.
The idea is to generalize the group theoretical formulation of SLE corresponding to WZW theory presented in our previous works\cite{SK2017,SK2018a},
in which the author considered a random process $\scr{G}_{t}$ on an infinite dimensional Lie group denoted by $\mrm{Aut}_{+}\mcal{O}\ltimes G_{+}(\mcal{O})$,
where $\mrm{Aut}_{+}\mcal{O}$ governs formal coordinate transformations at infinity,
$G$ is a finite dimensional simple complex Lie group and $G_{+}(\mcal{O})$ is a subgroup of the loop group for $G$.
The analogous object of this infinite dimensional Lie group in super WZW theory seems to be obvious,
it is expected to be obtained by replacing the finite dimensional Lie group $G$ by a finite dimensional Lie supergroup.
We will find, however, that the appropriate analogy forces us to add some Grassmann variables and take an infinite dimensional Lie group
denoted by $\mrm{Aut}_{+}\mcal{O}\ltimes G_{+}(\mcal{O}\otimes \bigwedge)$ as the target space of a random process $\scr{G}_{t}$.
In the previous works\cite{SK2017,SK2018a}, the author found that applying the random process $\scr{G}_{t}$ on a highest weight vector $\ket{v_{\Lambda}}$
of a certain highest weight representation of the corresponding affine Lie algebra,
the resulting random process $\scr{G}_{t}\ket{v_{\Lambda}}$ on (a formal completion of) the representation is a local martingale
with finely tuned parameters.
In the super case, since we extend the target group by Grassmann variables, correspondingly we have to take the tensor product of
a representation of an affine Lie superalgebra and a Grassmann algebra.
In Sect. \ref{sect:random_process}, we will see that for a highest weight vector $\ket{v_{\Lambda}}$, a potential local martingale is obtained by
\begin{equation}
	\int d\eta \scr{G}_{t}\ket{v_{\Lambda}}\otimes (1+\eta),
\end{equation}
where $\int d\eta$ is the Berezin integral over the Grassmann algebra.
After these observations, we can read off a set of SDEs from $\scr{G}_{t}$ whose solution is regarded as SLE with internal symmetry
described by an affine Lie superalgebra.
Table \ref{table:group_theoretical} shows the fundamental objects in the group theoretical formulation of SLE.

\begin{table}[h]
\caption{Fundamental objects in the group theoretical formulation of SLE}
{\begin{tabular}{|c|c|c|c|}\hline
	& Virasoro & WZW & super WZW \\ \hline
	Target group & $\mrm{Aut}_{+}\mcal{O}$ & $\mrm{Aut}_{+}\mcal{O}\ltimes G_{+}(\mcal{O})$ & $\mrm{Aut}_{+}\mcal{O}\ltimes G_{+}(\mcal{O}\otimes\bigwedge)$ \\ \hline
	Local martingale & $Q(\rho_{t})\ket{c,h}$ & $\scr{G}_{t}\ket{v_{\Lambda}}$ & $\int d\eta \scr{G}_{t}\ket{v_{\lambda}}\otimes (1+\eta)$ \\ \hline
\end{tabular}}
\label{table:group_theoretical}
\end{table}

This paper is organized as follows:
In Sect. \ref{sect:affine_Lie_super}, we recall the notion of affine Lie superalgebras and their representation theory.
In Sect. \ref{sect:internal_symmetry}, we introduce the infinite dimensional Lie group denoted by $\mrm{Aut}_{+}\mcal{O}\ltimes G_{+}(\mcal{O}\otimes\bigwedge)$ above.
A random process on the infinite dimensional Lie group is proposed in Sect. \ref{sect:random_process} as a generalization of the previous works \cite{SK2017,SK2018a}.
We focus our attention to the case that the underlying finite dimensional Lie superalgebra is $\mfrak{osp}(1|2)$ in Sect. \ref{sect:osp12}
and write down SDEs explicitly.
We also show a computation of local martingales associated with a solution from a representation of $\what{\mfrak{osp}(1|2)}$.
In Sect. \ref{sect:discussion}, we make some discussion on the result and directions of future research.

\section*{Acknowledgements}
The author is grateful to K. Sakai for leading him to this field of research and suggesting this topic,
and to R. Sato for teaching him much on Lie superalgabras.
This work was supported by a Grant-in-Aid for JSPS Fellows (Grant No. 17J09658).

\section{Affine Lie superalgebras and representations}
\label{sect:affine_Lie_super}
In this section, we recall the notion of affine Lie superalgebras and their representation theory.
We refer the readers to books\cite{ChengWang2013,Wakimoto2001} for a more detailed exposition.
Let $\mfrak{g}=\mfrak{g}_{\bar{0}}\oplus\mfrak{g}_{\bar{1}}$ be a finite dimensional Lie superalgebra of basic classical type except for $A(1|1)$,
and let $(\cdot|\cdot)$ be the nondegenerate invariant even supersymmetric bilinear form on $\mfrak{g}$
normalized so that the square norm of the highest root is 2.
The exception of $A(1|1)$ is not essential and is just for avoiding complexity of description
due to the fact that root spaces of $A(1|1)$ are not one-dimensional.
With these data, the corresponding affine Lie superalgebra is defined by
$\what{\mfrak{g}}=\mfrak{g}\otimes\mbb{C}[\zeta,\zeta^{-1}]\oplus\mbb{C}K$
with parity
\begin{align}
	\what{\mfrak{g}}_{\bar{0}}&=\mfrak{g}_{\bar{0}}\otimes\mbb{C}[\zeta,\zeta^{-1}]\oplus\mbb{C}K,&
	\what{\mfrak{g}}_{\bar{1}}&=\mfrak{g}_{\bar{1}}\otimes\mbb{C}[\zeta,\zeta^{-1}],
\end{align}
and Lie brackets
\begin{align*}
	[X(m),Y(n)]&=[X,Y](m+n)+m(X|Y)\delta_{m+n,0}K, & [K,\what{\mfrak{g}}]&=\{0\},
\end{align*}
where we denote $X\otimes \zeta^{m}$ for $X\in\mfrak{g}$ and $m\in\mbb{Z}$ by $X(m)$.

Let $M$ be a representation of $\mfrak{g}$,
then $M$ becomes a representation of $\mfrak{g}\otimes\mbb{C}[\zeta]\oplus\mbb{C}K$ so that
$\mfrak{g}\otimes \zeta^{0}$ acts naturally, $\mfrak{g}\otimes\mbb{C}[\zeta]\zeta$ acts trivially and
$K$ acts as multiplication by $k\in\mbb{C}$.
We extend this action to an action of the whole algebra by
\begin{equation}
	\what{M}_{k}=\mrm{Ind}_{\mfrak{g}\otimes\mbb{C}[\zeta]\oplus\mbb{C}K}^{\what{\mfrak{g}}}M=U(\what{\mfrak{g}})\otimes_{U(\mfrak{g}\otimes\mbb{C}[\zeta]\oplus\mbb{C}K)}M.
\end{equation}
By the Poinca\'{e}-Birkhoff-Witt theorem, $\what{M}_{k}$ is isomorphic to $U(\mfrak{g}\otimes\mbb{C}[\zeta^{-1}]\zeta^{-1})\otimes M$
as a vector space or a $U(\mfrak{g}\otimes\mbb{C}[\zeta^{-1}]\zeta^{-1})$-module.
The number $k$ introduced above is called the level of the representation.

We fix a Cartan subalgebra $\mfrak{h}\subset\mfrak{g}_{\bar{0}}$.
Let $L(\Lambda)$ be a finite dimensional highest weight representation of $\mfrak{g}$ of highest weight $\Lambda\in\mfrak{h}^{\ast}$.
Although $L(\Lambda)$ is irreducible, the induced representation $\what{L(\Lambda)}_{k}$ is not necessarily irreducible,
and in that case we denote its irreducible quotient by $L_{k}(\Lambda)$.

On a representation $L_{k}(\Lambda)$ of $\what{\mfrak{g}}$, we can define an action of the Virasoro algebra via the (super analogue of) Sugawara construction.
Let $\{X_{a}\}_{a=1}^{\dim\mfrak{g}}$ be a basis of $\mfrak{g}$, each element of which is homogeneous with respect to the $\mbb{Z}_{2}$-gradation,
and let $\{X^{a}\}_{a=1}^{\dim\mfrak{g}}$ be its dual basis with respect to $(\cdot|\cdot)$ that is characterized by $(X_{a}|X^{b})=\delta_{ab}$.
Then the operators
\begin{equation}
	\label{eq:Sugawara}
	L_{n}=\frac{1}{2(k+h^{\vee})}\sum_{a=1}^{\dim\mfrak{g}}\sum_{k\in\mbb{Z}}(-1)^{p(X_{a})}\no{X_{a}(n-k)X^{a}(k)},\ \ n\in\mbb{Z}
\end{equation}
define an action of the Virasoro algebra of central charge $c_{k}=\frac{k\mrm{sdim}\mfrak{g}}{k+h^{\vee}}$.
Here $h^{\vee}$ is the dual Coxeter number of $\mfrak{g}$, $p(X)$ denotes the $\mbb{Z}_{2}$-degree of $X$,
and the normal ordered product is defined by
\begin{equation}
	\no{A(p)B(q)}=
	\begin{cases}
		A(p)B(q), 	& p\le q,\\
		(-1)^{p(B)p(A)}B(q)A(p),	& p>q.
	\end{cases}
\end{equation}
The superdimension is defined by $\mrm{sdim}\mfrak{g}=\dim\mfrak{g}_{\bar{0}}-\dim\mfrak{g}_{\bar{1}}$.
Notice that the normal ordered product is necessary to make operators well-defined on a representation space $L_{k}(\Lambda)$
of an affine Lie superalgebra.
In our application to SLE, we need a more convenient expression of the Virasoro generators.
Let $\{J_{a}\}_{a=1}^{\dim\mfrak{g}_{\bar{0}}}$ be an orthonormal basis of the even part $\mfrak{g}_{\bar{0}}$ with respect to $(\cdot|\cdot)$.
We also denote the set of odd roots by $\Delta^{\mrm{odd}}$ and the set of positive odd roots by $\Delta_{+}^{\mrm{odd}}$.
For a basic classical Lie superalgebra $\mfrak{g}$ except for $A(1|1)$, each root space is one dimensional.
Then we normalize a basis $E_{\alpha}$ of a root space $\mfrak{g}_{\alpha}$ for $\alpha\in \Delta^{\mrm{odd}}$ by
$(E_{\alpha}|E_{-\alpha})=1$ for $\alpha\in\Delta_{+}^{\mrm{odd}}$.
The Virasoro generators in Eq.(\ref{eq:Sugawara}) are also expressed as
\begin{align}
	\label{eq:Sugawara_refined}
	L_{n}=\frac{1}{2(k+h^{\vee})}\sum_{k\in\mbb{Z}}\Biggl[&\sum_{a=1}^{\dim\mfrak{g}_{\bar{0}}}\no{J_{a}(n-k)J_{a}(k)} \\
	&+\sum_{\alpha\in\Delta_{+}^{\mrm{odd}}}\left(\no{E_{-\alpha}(n-k)E_{\alpha}(k)}-\no{E_{\alpha}(n-k)E_{-\alpha}(k)}\right)\Biggr].\notag
\end{align}
As has been already noted, the exception of $A(1|1)$ is just for simplification of description,
and we can treat $A(1|1)$ in a parallel way noticing multiplicity in root spaces.

\section{Target: infinite dimensional Lie group}
\label{sect:internal_symmetry}
In this section, we introduce an infinite dimensional Lie group, which will become the target space of our random process in Sect. \ref{sect:random_process}.

Firstly, we define an infinite dimensional Lie group that governs formal coordinate transformations following the book\cite{FrenkelBen-Zvi2004}.
Let $\mcal{O}=\mbb{C}[[z^{-1}]]$ be a complete topological $\mbb{C}$-algebra of formal power series
and $D=\mrm{Spec}\mcal{O}$ be the formal disc attached to infinity.
Here we regard $z$ as a coordinate at the origin.
Then we define an infinite dimensional Lie group $\mrm{Aut}\mcal{O}$ by one consisting of continuous automorphisms of $\mcal{O}$,
each element $\rho$ of which is identified with a formal power series
\begin{equation}
	\label{eq:formal_power_series}
	\rho(z)=a_{1}z+a_{0}+a_{-1}z^{-1}+\cdots \in z\mbb{C}[[z^{-1}]]
\end{equation}
such that $a_{1}\neq 0$.
The group law of $\mrm{Aut}\mcal{O}$ is defined by $(\rho\ast\mu)(z):=\mu(\rho(z))$ for $\rho,\mu\in\mrm{Aut}\mcal{O}$.
A significant subgroup $\mrm{Aut}_{+}\mcal{O}$ is defined by
\begin{equation}
	\mrm{Aut}_{+}\mcal{O}=\{z+a_{0}+a_{-1}z^{-1}+\cdots\}
\end{equation}
under the identification in Eq.(\ref{eq:formal_power_series}).
The Lie algebra of $\mrm{Aut}\mcal{O}$ consists of holomorphic vector fields at infinity and is realized as $\mrm{Der}_{0}\mcal{O}=z\mbb{C}[[z^{-1}]]\del_{z}$.
Correspondingly, the Lie algebra of $\mrm{Aut}_{+}\mcal{O}$ is identified with $\mrm{Der}_{+}\mcal{O}=\mbb{C}[[z^{-1}]]\del_{z}$.
These Lie algebras are roughly regarded as ``halfs" of the Virasoro algebra, and thus seem to act on a representation space of the Virasoro algebra,
but to verify this action, we have to ``complete" the representation space.
In Sect. \ref{sect:affine_Lie_super}, we saw that on a highest weight representation $L_{k}(\Lambda)$ of an affine Lie superalgebra
one can define an action of the Virasoro algebra via the Sugawara construction.
It is easily shown that this space is diagonalized by the action of $L_{0}$ so that $L_{k}(\Lambda)=\bigoplus_{n\in \mbb{Z}_{\ge 0}}L_{k}(\Lambda)_{h_{\Lambda}+n}$,
where $L_{k}(\Lambda)_{h}=\{v\in L_{k}(\Lambda)|L_{0}v=hv\}$ is an eigenspace of $L_{0}$.
The eigenvalue of the ``top component" is given by $h_{\Lambda}=\frac{(\Lambda|\Lambda+2\rho)}{2(k+h^{\vee})}$ with $\rho$ being the Weyl vector.
Then the formal completion of $L_{k}(\Lambda)$ is defined by $\overline{L_{k}(\Lambda)}=\prod_{n\in\mbb{Z}_{\ge 0}}L_{k}(\Lambda)_{h_{\Lambda}+n}$,
on which the Lie algebras $\mrm{Der}_{0}\mcal{O}$ and $\mrm{Der}_{+}\mcal{O}$ act by the assignment $-z^{n+1}\del_{z}\mapsto L_{n}$.
The action of the Lie algebra $\mrm{Der}_{+}\mcal{O}$ on $\overline{L_{k}(\Lambda)}$ can be exponentiated to an action of $\mrm{Aut}_{+}\mcal{O}$.
If $h_{\Lambda}$ is an integer, the action of $\mrm{Der}_{0}\mcal{O}$ also can be exponentiated to give an action of $\mrm{Aut}\mcal{O}$,
which is, however, not necessary in the present paper.
A detailed description of this action can be seen in the previous paper\cite{SK2018a}.
We denote this action of $\mrm{Aut}_{+}\mcal{O}$ by $Q:\mrm{Aut}_{+}\mcal{O}\to\mrm{Aut}(\overline{L_{k}(\Lambda)})$.

We next introduce a group of internal symmetry.
Since we are considering super WZW theory, the internal symmetry is described by a Lie supergroup,
which is abstractly defined as a group object in the category of supermanifolds \cite{Varadarajan2004}.
This definition is, however, not convenient for our purpose to construct SLE.
Instead we directly treat a set of super-algebra-valued points of a Lie supergroup.
To present an idea, we let $\mfrak{k}=\mfrak{k}_{\bar{0}}\oplus\mfrak{k}_{\bar{1}}$ be a Lie superalgebra
and $\bigwedge=\bigwedge[\eta_{1},\cdots,\eta_{n}]$ be a Grassmann algebra generated by $n$ Grassmann variables $\eta_{1},\cdots,\eta_{n}$.
The Grassmann envelop \cite{BerezinTolstoy1981, FrappatSciarrinoSorba2000} of $\mfrak{k}$ by $\bigwedge$ is the even part of $\mfrak{k}\otimes \bigwedge$, or explicitly
\begin{equation}
	\mfrak{k}(\bigwedge)=(\mfrak{k}_{\bar{0}}\otimes \bigwedge{}_{\bar{0}})\oplus (\mfrak{k}_{\bar{1}}\otimes\bigwedge{}_{\bar{1}}).
\end{equation}
It is a (non-super) Lie algebra with Lie bracket defined by $[X\otimes \eta,Y\otimes \pr{\eta}]=[X,Y]\otimes \eta\pr{\eta}$.
We formally exponentiate each element of the Grassmann envelop $\mfrak{k}(\bigwedge)$,
and denote a Lie group generated by such objects by $K(\bigwedge)=\braket{e^{X\otimes\eta}}$.
Let us apply this construction to our case.

We take $\mfrak{k}=\mfrak{g}\otimes\mbb{C}[[\zeta^{-1}]]\zeta^{-1}$ and $\bigwedge=\bigwedge[\eta_{1},\eta_{2}]$.
Then we denote the resulting Lie group $K(\bigwedge)$ by $G_{+}(\mcal{O}\otimes \bigwedge)$.
The spirit of this notation comes from the observation that we can regard this Lie group as a set of ($\mcal{O}\otimes\bigwedge$)-valued points
of a Lie supergroup $G$ whose Lie superalgebra is the underlying one $\mfrak{g}$.
To describe an example of the group law for $G_{+}(\mcal{O}\otimes\bigwedge)$, let us recall the Campbell-Baker-Hausdorff formula:
\begin{equation}
	\exp(A)\exp(B)=\exp\left(A+B+\frac{1}{2}[A,B]+\frac{1}{12}\left([A,[A,B]]+[B,[B,A]]\right)+\cdots\right)
\end{equation}
for not necessarily commutative symbols $A$ and $B$.
Let us apply this to the case of $A=X(\zeta)\otimes \eta_{1}$, $B=Y(\zeta)\otimes \eta_{2}$
for $X(\zeta), Y(\zeta)\in\mfrak{g}_{\bar{1}}\otimes\mbb{C}[[\zeta^{-1}]]\zeta^{-1}$.
Then they are exponentiated to give elements $\exp(X(\zeta)\otimes \eta_{1})$ and $\exp(Y(\zeta)\otimes \eta_{2})$
of the group $G_{+}(\mcal{O}\otimes\bigwedge)$ and the product among them is identified with
\begin{align}
	&\exp(X(\zeta)\otimes\eta_{1})\exp(Y(\zeta)\otimes \eta_{2}) \\
	&=\exp\left(X(\zeta)\otimes\eta_{1}+Y(\zeta)\otimes \eta_{2}+\frac{1}{2}[X(\zeta),Y(\zeta)]\otimes\eta_{1}\eta_{2}\right),\notag
\end{align}
due to the nilpotency of Grassmann variables.
Note that the right hand side defines an element of the group $G_{+}(\mcal{O}\otimes\bigwedge)$.

We have a natural action of $\mrm{Aut}_{+}\mcal{O}$ on the group $G_{+}(\mcal{O}\otimes\bigwedge)$
by transformation of the loop variable $\zeta$
so that $\rho\in\mrm{Aut}_{+}\mcal{O}$ transforms $\exp(X(\zeta)\otimes \eta)$ to $\exp(X(\rho(\zeta))\otimes \eta)$.
Thus we have a semi-direct product group $\mrm{Aut}_{+}\mcal{O}\ltimes G_{+}(\mcal{O}\otimes\bigwedge)$,
which will be the target group of a random process considered in Sect. \ref{sect:random_process}.
Intuitively, this semi-direct product can be considered as one generated by elements in $\mrm{Aut}_{+}\mcal{O}$ and $G_{+}(\mcal{O}\otimes\bigwedge)$
with the following relations being imposed:
\begin{equation}
	\label{eq:semi-direct_product}
	\rho \exp(X(\zeta)\otimes \eta)\rho^{-1}=\exp(X(\rho(\zeta))\otimes \eta)
\end{equation}
for $\rho\in \mrm{Aut}_{+}\mcal{O}$ and $\exp(X(\zeta)\otimes \eta)\in G_{+}(\mcal{O}\otimes\bigwedge)$.

Finally in this section, we see a representation of this infinite dimensional Lie group $\mrm{Aut}_{+}\mcal{O}\ltimes G_{+}(\mcal{O}\otimes\bigwedge)$.
As we have already seen, the group of coordinate transformations $\mrm{Aut}_{+}\mcal{O}$ acts on the formal completion $\overline{L_{k}(\Lambda)}$
of a representation of the affine Lie superalgebra.
Since the group of internal symmetry involves Grassmann variables, it cannot act on the same space, but acts on $\overline{L_{k}(\Lambda)}\otimes \bigwedge$ in a natural way.
Moreover since the action of the Virasoro algebra on $L_{k}(\Lambda)$ is defined by the Sugawara construction,
it is compatible with the semi-direct product structure, which verifies the action of the group $\mrm{Aut}_{+}\mcal{O}\ltimes G_{+}(\mcal{O}\otimes\bigwedge)$
on a space $\overline{L_{k}(\Lambda)}\otimes \bigwedge$.
For example, an element in $\mrm{Aut}_{+}\mcal{O}$ is represented by an operator like $\exp\left(\sum_{j<0}v_{j}L_{j}\right)$ for $v_{j}\in \mbb{C}$,
which just acts on $\overline{L_{k}(\Lambda)}$.
As another example, an element $\exp(X(\zeta)\otimes \eta_{1})$ in $G_{+}(\mcal{O}\otimes\bigwedge)$
for $X(\zeta)\in \mfrak{g}_{\bar{1}}\otimes\mbb{C}[[\zeta^{-1}]]\zeta^{-1}$
acts on a vector $v\otimes 1\in \overline{L_{k}(\Lambda)}\otimes \bigwedge$ as
\begin{equation}
	\exp(X(\zeta)\otimes \eta_{1})(v\otimes 1)=v\otimes1 +(X(\zeta)v)\otimes \eta_{1}.
\end{equation}
One can observe that the right hand side is well-defined in $\overline{L_{k}(\Lambda)}\otimes\bigwedge$.

\section{Random process on a symmetry group}
\label{sect:random_process}

In the previous Sect. \ref{sect:internal_symmetry}, we defined an infinite dimensional Lie group denoted by $\mrm{Aut}_{+}\mcal{O}\ltimes G_{+}(\mcal{O}\otimes\bigwedge)$,
which governs coordinate transformation and internal symmetry.
In this section, we introduce a random process on the infinite dimensional Lie group that induces a generalization of SLE
that possesses internal symmetry described by an affine Lie superalgebra.
Our construction is a natural generalization of one presented in the previous works\cite{SK2017,SK2018a} for WZW theory.

We begin with a simple observation.
Let $\ket{0}$ be the highest weight vector of $L_{k}(0)$, which is called the vacuum.
From the concrete description of the Virasoro generators in Eq.(\ref{eq:Sugawara_refined}),
we have
\begin{align}
	L_{-2}\ket{0}=\frac{1}{2(k+h^{\vee})}\Biggl(&\sum_{a=1}^{\dim\mfrak{g}_{\bar{0}}}J_{a}(-1)^{2} \\
	&+\sum_{\alpha\in\Delta_{+}^{\mrm{odd}}}\left(E_{-\alpha}(-1)E_{\alpha}(-1)-E_{\alpha}(-1)E_{-\alpha}(-1)\right)\Biggr)\ket{0}.\notag
\end{align}
From this and the fact that the vacuum vector is translation invariant: $L_{-1}\ket{0}=0$, we have
\begin{align}
	\Biggl[&-2L_{-2}+\frac{\kappa}{2}L_{-1}^{2}  \\
	&+\frac{\tau}{2}\left(\sum_{a=1}^{\dim\mfrak{g}_{\bar{0}}}J_{a}(-1)^{2}
	+\sum_{\alpha\in\Delta_{+}^{\mrm{odd}}}\left(E_{-\alpha}(-1)E_{\alpha}(-1)-E_{\alpha}(-1)E_{-\alpha}(-1)\right)\right)\Biggr]\ket{0}=0 \notag
\end{align}
for arbitrary $\kappa$ and $\tau=\frac{2}{k+h^{\vee}}$.
As a generalization of this, given level $k$, we assume that we can choose positive numbers $\kappa$ and $\tau$ so that
\begin{align}
	\label{eq:annihilator}
	\Biggl[&-2L_{-2}+\frac{\kappa}{2}L_{-1}^{2}  \\
	&+\frac{\tau}{2}\left(\sum_{a=1}^{\dim\mfrak{g}_{\bar{0}}}J_{a}(-1)^{2}
	+\sum_{\alpha\in\Delta_{+}^{\mrm{odd}}}\left(E_{-\alpha}(-1)E_{\alpha}(-1)-E_{\alpha}(-1)E_{-\alpha}(-1)\right)\right)\Biggr]\ket{v_{\Lambda}}=0 \notag
\end{align}
where $\ket{v_{\Lambda}}$ is the highest weight vector of $L_{k}(\Lambda)$.
The positivity of the parameters $\kappa$ and $\tau$ is required since
they later play roles of variances of Brownian motions.
This annihilating operator of a highest weight vector motivates us to consider a random process $\scr{G}_{t}$ on 
the infinite dimensional Lie group $\mrm{Aut}_{+}\mcal{O}\ltimes G_{+}(\mcal{O}\otimes\bigwedge)$ that satisfies the following SDE:
\begin{align}
	\label{eq:random_process_on_group}
	\scr{G}_{t}^{-1}d\scr{G}_{t}
	=&\Biggl[-2L_{-2}+\frac{\kappa}{2}L_{-1}^{2}  \\
	&+\frac{\tau}{2}\Biggl(\sum_{a=1}^{\dim\mfrak{g}_{\bar{0}}}J_{a}(-1)^{2} \notag \\
	&\hspace{30pt}+\eta_{1}\eta_{2}\sum_{\alpha\in\Delta_{+}^{\mrm{odd}}}\left(E_{-\alpha}(-1)E_{\alpha}(-1)-E_{\alpha}(-1)E_{-\alpha}(-1)\right)\Biggr)\Biggr] dt \notag \\
	&+L_{-1}dB_{t}^{(0)}+\sum_{a=1}^{\dim\mfrak{g}_{\bar{0}}}J_{a}(-1)dB_{t}^{(a)}\notag \\
	&+\sum_{\alpha\in\Delta_{+}^{\mrm{odd}}}(\eta_{1}E_{-\alpha}(-1)+\eta_{2}E_{\alpha}(-1))dB_{t}^{(\alpha)}. \notag
\end{align}
Here $B_{t}^{(0)}$, $\{B_{t}^{(a)}\}_{a=1}^{\dim\mfrak{g}_{\bar{0}}}$ and $\{B_{t}^{(\alpha)}\}_{\alpha\in\Delta_{+}^{\mrm{odd}}}$ are
mutually independent Brownian motions whose covariance are $\kappa$ for $B_{t}^{(0)}$ and $\tau$ for the others.
Note that for Grassmann variables $\eta_{1}$ and $\eta_{2}$, we have
\begin{equation}
	(\eta_{1}E_{-\alpha}(-1)+\eta_{2}E_{\alpha}(-1))^{2}=\eta_{1}\eta_{2}(E_{-\alpha}(-1)E_{\alpha}(-1)-E_{\alpha}(-1)E_{-\alpha}(-1)),
\end{equation}
thus the SDE in Eq.(\ref{eq:random_process_on_group}) is in the standard form of one for a random process on a Lie group.
We write the operator in front of $dt$ in Eq.(\ref{eq:random_process_on_group}) as $\Xi(\kappa,\tau)$,
which acts on $L_{k}(\Lambda)\otimes \bigwedge$.
Now we observe that
\begin{align}
	&\int d\eta_{2}d\eta_{1}\Xi(\kappa,\tau)\ket{v_{\Lambda}}\otimes (1+\eta_{1}\eta_{2}) \\
	&=\Biggl[-2L_{-2}+\frac{\kappa}{2}L_{-1}^{2} \notag \\
	&\hspace{20pt}+\frac{\tau}{2}\left(\sum_{a=1}^{\dim\mfrak{g}_{\bar{0}}}J_{a}(-1)^{2}
	+\sum_{\alpha\in\Delta_{+}^{\mrm{odd}}}\left(E_{-\alpha}(-1)E_{\alpha}(-1)-E_{\alpha}(-1)E_{-\alpha}(-1)\right)\right)\Biggr]\ket{v_{\Lambda}} \notag \\
	&=0,\notag
\end{align}
where $\int d\eta_{2}d\eta_{1}$ is the Berezin integral over the Grassmann algebra $\bigwedge$ and
the last equality is due to the assumption in Eq.(\ref{eq:annihilator}).
This implies that the random process
\begin{equation}
	\label{eq:local_martingale}
	\int d\eta_{2}d\eta_{1}\scr{G}_{t}\ket{v_{\Lambda}}\otimes (1+\eta_{1}\eta_{2})
\end{equation}
is a local martingale in $\overline{L_{k}(\Lambda)}$.

To see that the random process $\scr{G}_{t}$ on $\mrm{Aut}_{+}\mcal{O}\ltimes G_{+}(\mcal{O}\otimes \bigwedge)$
actually induces SLE with internal degrees of freedom, we factorize it so that
\begin{equation}
	\scr{G}_{t}=\Theta_{t}Q(\rho_{t}),
\end{equation}
where $\rho_{t}$ and $\Theta_{t}$ are random processes on $\mrm{Aut}_{+}\mcal{O}$ and $G_{+}(\mcal{O}\otimes\bigwedge)$, respectively.
Under this ansatz, we can find with help of Eq.(\ref{eq:semi-direct_product}) that
they satisfy the following SDEs:
\begin{align}
	\label{eq:geometric_operator}
	Q(\rho_{t})^{-1}dQ(\rho_{t})=&\left(-2L_{-2}+\frac{\kappa}{2}L_{-1}^{2}\right)dt +L_{-1}dB_{t}^{(0)}, \\
	\label{eq:algebraic}
	\Theta_{t}^{-1}d\Theta_{t}=&\frac{\tau}{2}\Biggl[\sum_{a=1}^{\dim\mfrak{g}_{\bar{0}}}(J_{a}\otimes \rho_{t}(\zeta)^{-1})^{2}  \\
	&\hspace{15pt}+\eta_{1}\eta_{2}\sum_{\alpha\in\Delta_{+}^{\mrm{odd}}}\Bigl(E_{-\alpha}\otimes\rho_{t}(\zeta)^{-1}E_{\alpha}\otimes\rho_{t}(\zeta)^{-1}\notag \\
	&\hspace{80pt}-E_{\alpha}\otimes\rho_{t}(\zeta)^{-1}E_{-\alpha}\otimes\rho_{t}(\zeta)^{-1}\Bigr)\Biggr]dt \notag \\
	&+\sum_{a=1}^{\dim\mfrak{g}_{\bar{0}}}J_{a}\otimes\rho_{t}(\zeta)^{-1}dB_{t}^{(a)} \notag \\
	&+\sum_{\alpha\in\Delta_{+}^{\mrm{odd}}}\left(\eta_{1}E_{-\alpha}\otimes\rho_{t}(\zeta)^{-1}+\eta_{2}E_{\alpha}\otimes\rho_{t}(\zeta)^{-1}\right) dB_{t}^{(\alpha)}.\notag
\end{align}
Since the action of the group $\mrm{Aut}_{+}\mcal{O}$ on $\overline{L_{k}(\Lambda)}$ is defined by assigning
an infinitesimal transformation $\ell_{n}=-z^{n+1}\del_{z}$ to the Virasoro generator $L_{n}$,
the first equation (\ref{eq:geometric_operator}) gives an SDE for $\rho_{t}$ so that
\begin{equation}
	\rho_{t}^{-1}d\rho_{t}=\left(-2\ell_{-2}+\frac{\kappa}{2}\ell_{-1}^{2}\right)dt+\ell_{-1}dB_{t}^{(0)}.
\end{equation}
When we apply both sides to the variable $z$, we find that the value $\rho_{t}(z)$ satisfies the following SDE:
\begin{equation}
	\label{eq:geometric}
	d\rho_{t}(z)=\frac{2}{\rho_{t}(z)}dt-dB_{t}^{(0)},
\end{equation}
of which a solution is equivalent to SLE characterized by the parameter $\kappa$ by setting $g_{t}(z)=\rho_{t}(z)+B_{t}^{(0)}$.
The SDE (\ref{eq:algebraic}) is also a generalization of one on internal degrees of freedom 
of SLE corresponding to WZW theory \cite{SK2017,SK2018a}.
Thus we can regard a solution of the set of SDEs (\ref{eq:geometric}) and (\ref{eq:algebraic}) as
SLE with internal symmetry described by an affine Lie superalgebra,
and can compute local martingales associated with the solution from a representation-space-valued one in Eq.(\ref{eq:local_martingale}).
Note that this structure is prototypical in SLE/CFT correspondence\cite{BauerBernard2002, BauerBernard2003a,BauerBernard2003b,Rasmussen2004a, NagiRasmussen2005,Rasmussen2007,BettelheimGruzbergLudwigWiegmann2005,AlekseevBytskoIzyurov2011}.
Let us remark that the parameters $\kappa$ and $\tau$ appearing in the annihilating operator (\ref{eq:annihilator})
correspond to variances of Brownian motions,
and thus in particular, $\kappa$ characterizes the geometry (e.g. fractal dimension) of SLE traces.
For example, each SLE trace is almost surely a simple path if $\kappa\le 4$.
The other parameter $\tau$ also characterizes the behavior of the random process along the internal degrees of freedom,
while its geometric interpretation such as how the shape of sample paths in the internal space depends on $\tau$ is unclear so far.

\section{Example}
\label{sect:osp12}
In this section, we construct a random process $\scr{G}_{t}$ introduced in previous Sect.\ref{sect:random_process}
in a concrete way in case that $\mfrak{g}=\mfrak{osp}(1|2)$,
and show a computation of local martingale associated with the solution.

The Lie superalgebra $\mfrak{osp}(1|2)$ is defined by
\begin{align}
	\mfrak{osp}(1|2)_{\bar{0}}&=\mfrak{sl}_{2}=\mbb{C}E\oplus \mbb{C}H \oplus \mbb{C}F, &
	\mfrak{osp}(1|2)_{\bar{1}}&=\mbb{C}e\oplus\mbb{C}f,
\end{align}
with the Lie superbracket being given by
\begin{align*}
	[H,e]&=e, & [H,f]&=-f, &&\\
	[E,f]&=-e, & [F,e]&=-f, & [E,e]&=[F,f]=0,\\
	[e,e]&=E, & [f,f]&=-F, & [e,f]&=H.
\end{align*}
The even part is isomorphic to $\mfrak{sl}_{2}$ as a Lie algebra and we omitted the Lie bracket among even elements.
The nondegenerate even invariant supersymmetric bilinear form $(\cdot|\cdot)$ coincides with one of $\mfrak{sl}_{2}$ on the even part
and is characterized by $(e|f)=2$.
We fix a Cartan subalgebra $\mfrak{h}=\mbb{C}H$.
The set of roots is $\{\pm\alpha,\pm\frac{\alpha}{2}\}$, where $\pm\alpha$ are even roots and $\pm\frac{\alpha}{2}$ are odd roots.
Together with the normalization with respect to $(\cdot|\cdot)$, we set $E_{\frac{\alpha}{2}}=\frac{1}{\sqrt{2}}e$ and $E_{-\frac{\alpha}{2}}=\frac{1}{\sqrt{2}}f$.
Then the Sugawara construction in Eq.(\ref{eq:Sugawara_refined}) reads
\begin{align}
	L_{n}=\frac{1}{2(k+h^{\vee})}\sum_{k\in\mbb{Z}}\Biggl[ &\frac{1}{2}\no{H(n-k)H(k)}+\no{E(n-k)F(k)}+\no{F(n-k)E(k)}  \\
	&+\frac{1}{2}\left(\no{f(n-k)e(k)}-\no{e(n-k)f(k)}\right)\Biggr] \notag
\end{align}

We shall construct the random process $\Theta_{t}$ in $G_{+}(\mcal{O}\otimes \bigwedge)$ satisfying Eq.(\ref{eq:algebraic}) in a concrete way.
We take an orthonormal basis $\{X_{a}\}_{a=1}^{3}$ of the even part by
\begin{align}
	J_{1}&=\frac{1}{\sqrt{2}}H, & J_{2}&=\frac{1}{\sqrt{2}}(E+F), & J_{3}&=\frac{i}{\sqrt{2}}(E-F).
\end{align}
We set an ansatz $\Theta_{t}=\Theta^{\bar{1}}_{t}\Theta^{\bar{0}}_{t}$, with
\begin{align}
	\Theta^{\bar{0}}_{t}&=e^{E\otimes x^{E}_{t}(\zeta)}e^{H\otimes x^{H}_{t}(\zeta)}e^{F\otimes x^{F}_{t}(\zeta)}, \\
	\Theta^{\bar{1}}_{t}&=e^{\eta_{1}L^{1}_{t}}e^{\eta_{2}L^{2}_{t}}e^{\eta_{1}\eta_{2}L^{12}_{t}},
\end{align}
where $x^{E}_{t}(\zeta), x^{H}_{t}(\zeta), x^{F}_{t}(\zeta)$ are $\mbb{C}[[\zeta^{-1}]]\zeta^{-1}$-valued random processes,
$L^{1}_{t}, L^{2}_{t}$ are $\mfrak{g}_{\bar{1}}\otimes\mbb{C}[[\zeta^{-1}]]\zeta^{-1}$-valued random processes
and $L^{12}_{t}$ is a $\mfrak{g}_{\bar{0}}\otimes\mbb{C}[[\zeta^{-1}]]\zeta^{-1}$-valued random process.
Under this ansatz, the ``even" part $\Theta^{\bar{0}}_{t}$ satisfies
\begin{equation}
	(\Theta^{\bar{0}}_{t})^{-1}d\Theta^{\bar{0}}_{t}
	=\frac{\tau}{2}\sum_{a=1}^{\dim\mfrak{g}_{\bar{0}}}(J_{a}\otimes \rho_{t}(\zeta)^{-1})^{2}+\sum_{a=1}^{\dim\mfrak{g}_{\bar{0}}}J_{a}\otimes\rho_{t}(\zeta)^{-1}dB_{t}^{(a)},
\end{equation}
which coincides with the SDE on the internal degrees of freedom corresponding to $\what{\mfrak{sl}}_{2}$ investigated in the previous paper\cite{SK2018a}.
There the random processes $x^{E}_{t}(\zeta), x^{H}_{t}(\zeta), x^{F}_{t}(\zeta)$ have been shown to satisfy\cite{SK2018a}
\begin{align}
	dx^{E}_{t}(\zeta)=&-\frac{e^{2x^{H}_{t}(\zeta)}}{\sqrt{2}\rho_{t}(\zeta)}dB_{t}^{(2)}-\frac{ie^{2x^{H}_{t}(\zeta)}}{\sqrt{2}\rho_{t}(\zeta)}dB_{t}^{(3)}, \\
	dx^{H}_{t}(\zeta)=&-\frac{\tau}{2\rho_{t}(\zeta)^{2}}dt
				-\frac{1}{\sqrt{2}\rho_{t}(\zeta)}dB_{t}^{(1)}+\frac{x^{F}_{t}(\zeta)}{\sqrt{2}\rho_{t}(\zeta)}dB_{t}^{(2)}+\frac{ix^{F}_{t}(\zeta)}{\sqrt{2}\rho_{t}(\zeta)}dB_{t}^{(3)}, \\
	dx^{F}_{t}(\zeta)=&-\frac{\sqrt{2}x^{F}_{t}(\zeta)}{\rho_{t}(\zeta)}dB_{t}^{(1)}-\frac{1-x^{F}_{t}(\zeta)^{2}}{\sqrt{2}\rho_{t}(\zeta)}dB_{t}^{(2)}+\frac{i(1+x^{F}_{t}(\zeta)^{2})}{\sqrt{2}\rho_{t}(\zeta)}.
\end{align}

To describe the random processes $L^{1}_{t}$, $L^{2}_{t}$ and $L^{12}_{t}$, we write them as
\begin{align}
	L^{1}_{t}&=e\otimes x^{1,e}_{t}(\zeta)+f\otimes x^{1,f}_{t}(\zeta), \\
	L^{2}_{t}&=e\otimes x^{2,e}_{t}(\zeta)+f\otimes x^{2,f}_{t}(\zeta), \\
	L^{12}_{t}&=E\otimes x^{12,E}_{t}(\zeta)+H\otimes x^{12,H}_{t}(\zeta)+F\otimes x^{12,F}_{t}(\zeta),
\end{align}
where $x^{i,\natural}_{t}(\zeta)$ for $i=1,2$, $\natural=e,f$ and $x^{12,\natural}_{t}(\zeta)$ for $\natural=E,H,F$ are
$\mbb{C}[[\zeta^{-1}]]\zeta^{-1}$-valued random processes, the SDEs for which are to be derived.
This task is carried out by substituting these ansatz into Eq.(\ref{eq:algebraic}), and as a result, we have
\begin{align}
	dx^{1,e}_{t}(\zeta)=&\sqrt{2}(e^{x^{H}_{t}(\zeta)}+e^{-x^{H}_{t}(\zeta)}x^{E}_{t}(\zeta)x^{F}_{t}(\zeta))\rho_{t}(\zeta)^{-1}dB_{t}^{(\alpha/2)}, \\
	dx^{1,f}_{t}(\zeta)=&-\sqrt{2}e^{-x^{H}_{t}(\zeta)}x^{F}_{t}(\zeta)\rho_{t}(\zeta)^{-1}dB_{t}^{(\alpha/2)},\\
	dx^{2,e}_{t}(\zeta)=&-\sqrt{2}e^{-x^{H}_{t}(\zeta)}x^{E}_{t}(\zeta)\rho_{t}(\zeta)^{-1}dB_{t}^{(\alpha/2)}, \\
	dx^{2,f}_{t}(\zeta)=&\sqrt{2}e^{-x^{H}_{t}(\zeta)}\rho_{t}(\zeta)^{-1}dB_{t}^{(\alpha/2)}, \\
	dx^{12,E}_{t}(\zeta)=&\tau(1+e^{-2x^{H}_{t}(\zeta)}x^{E}_{t}(\zeta)x^{F}_{t}(\zeta))x^{E}_{t}(\zeta)\rho_{t}(\zeta)^{-2}dt \\
		&-\sqrt{2}x^{2,e}_{t}(\zeta)(e^{x^{H}_{t}(\zeta)}+e^{-x^{H}_{t}(\zeta)}x^{E}_{t}(\zeta)x^{F}_{t}(\zeta))\rho_{t}(\zeta)^{-1}dB_{t}^{(\alpha/2)}, \notag \\
	dx^{12,H}_{t}(\zeta)=&-\frac{\tau}{2}(1+2e^{-2x^{H}_{t}(\zeta)}x^{E}_{t}(\zeta)x^{F}_{t}(\zeta))\rho_{t}(\zeta)^{-2}dt  \\
		&+\sqrt{2}(x^{2,e}_{t}(\zeta)e^{-x^{H}_{t}(\zeta)}x^{F}_{t}(\zeta)\rho_{t}(\zeta)^{-1}  \notag \\
		&\hspace{30pt}-x^{2,f}_{t}(\zeta)(e^{x^{H}_{t}(\zeta)}+e^{-x^{H}_{t}(\zeta)}x^{E}_{t}(\zeta)x^{F}_{t}(\zeta)\rho_{t}(\zeta)^{-1}))dB_{t}^{(\alpha/2)}.\notag \\
	dx^{12,F}_{t}(\zeta)=&-\tau e^{-2x^{H}_{t}(\zeta)}x^{F}_{t}(\zeta)\rho_{t}(\zeta)^{-2}dt  \\
		&-\sqrt{2}x^{2,f}_{t}(\zeta)e^{-x^{H}_{t}(\zeta)}x^{F}_{t}(\zeta)\rho_{t}(\zeta)^{-1}dB_{t}^{(\alpha/2)}. \notag
\end{align}

Let us investigate when the annihilating condition Eq.(\ref{eq:annihilator}) holds for $\mfrak{g}=\mfrak{osp}(1|2)$.
We already know that the vacuum vector $\ket{0}$ gives an example, thus we seek one in a higher spin representation,
but we will see that there is no example of an annihilating operator of the form of Eq.(\ref{eq:annihilator}) on representations other than the vacuum representation
for $\mfrak{g}=\mfrak{osp}(1|2)$.
We set
\begin{equation}
	\psi=\left(-2L_{-1}+\frac{\kappa}{2}L_{-1}^{2}+\frac{\tau}{2}\sum_{a=1}^{\dim\mfrak{g}}(-1)^{p(X_{a})}X_{a}(-1)X^{a}(-1)\right)\ket{v_{\Lambda}},
\end{equation}
which is a candidate for a null vector.
The vector $\psi$ is a null vector if and only if $X(1)\psi=0$ and $X(2)\psi=0$ for arbitrary $X\in\mfrak{g}$.
These conditions are equivalent to
\begin{align}
	\label{eq:null_vector_condition_1}
	&\Biggl((\tau k- \tau h^{\vee}-2)X(-1)+\kappa X(0)L_{-1}  \\
	&\hspace{20pt}+\tau\sum_{a=1}^{\dim\mfrak{g}}(-1)^{p(X_{a})}[X,X_{a}](0)X^{a}(-1)\Biggr)\ket{v_{\Lambda}}=0, \notag \\
	\label{eq:null_vector_condition_2}
	&(\kappa+\tau h^{\vee}-4)X(0)\ket{v_{\Lambda}}=0.
\end{align}
Other than the vacuum representation $\Lambda=0$, the second condition Eq.(\ref{eq:null_vector_condition_2}) imposes $\kappa+\tau h^{\vee}-4=0$.
We can see that for $\mfrak{g}=\mfrak{osp}(1|2)$, the first condition Eq.(\ref{eq:null_vector_condition_1}) does not hold.
Indeed, for $X=E$, Eq.(\ref{eq:null_vector_condition_1}) yields
\begin{equation}
	\left((\tau k-\tau h^{\vee}-2+\tau(4+\lambda))E(-1)+\frac{\tau}{2}H(-1)\right)\ket{v_{\Lambda}}=0,
\end{equation}
where $\lambda$ is defined by $H(0)\ket{v_{\Lambda}}=\lambda \ket{v_{\Lambda}}$,
but it does not hold unless $\tau=0$.

As we have seen, an annihilating operator of the form in Eq.(\ref{eq:annihilator}) exists for the vacuum representation.
Then for arbitrary $\kappa>0$ and $\tau=\frac{2}{k+3/2}$, the quantity in Eq.(\ref{eq:local_martingale}) is a $\overline{L_{k}(0)}$-valued local martingale.
We can obtain infinitely many local martingales by taking the inner product among the representation-space-valued local martingale and vectors in $L_{k}(0)$.
Now we show an example: for an current field $E(z)=\sum_{n\in\mbb{Z}}E(n)z^{-n-1}$,
\begin{align}
	&\int d\eta_{2}d\eta_{1}\braket{0|E(z)\scr{G}_{t}|0}\otimes (1+\eta_{1}\eta_{2})  \\
	&=k(1+x^{1,f}_{t}(z)x^{2,e}_{t}(z)+2x^{12,H}_{t}(z))\del x^{F}_{t}(z) \notag \\
	&\hspace{10pt}-k(x^{1,f}(z)x^{2,f}_{t}(z)-x^{12,F}_{t}(z))(2\del x^{H}_{t}(z)+2e^{-2x^{H}_{t}(z)}x^{E}_{t}(z)\del x^{F}_{t}(z)) \notag\\
	&\hspace{10pt}-2kx^{1,f}_{t}(z)\del x^{2,f}_{t}(z)+k\del x^{12,F}_{t}(z) \notag
\end{align}
is a local martingale, while its interpretation is not clear so far.

\section{Discussion}
\label{sect:discussion}
In this paper, we consider a generalization of SLE that corresponds to super WZW theory.
Our present work is a generalization of previous works \cite{SK2017,SK2018a},
which formulated SLE corresponding to WZW theory appearing in the literature\cite{BettelheimGruzbergLudwigWiegmann2005,AlekseevBytskoIzyurov2011} along the line of the group theoretical formulation\cite{BauerBernard2003a,Rasmussen2007}.
We began with a random process on an infinite dimensional Lie group that governs coordinate transformations and internal symmetry
and saw that it induces a set of SDEs whose solution is regarded as SLE with internal symmetry described by an affine Lie superalgebra.
Our construction also allows one to compute local martingales associated with the solution from a single representation-space-valued local martingale in Eq.(\ref{eq:local_martingale})
provided an annihilating condition Eq.(\ref{eq:annihilator}).

Let us make some discussion on further direction of research.
A significant problem is to find an example of an annihilating operator of a highest weight vector of the form in Eq.(\ref{eq:annihilator}).
It was shown in this paper that for $\mfrak{g}=\mfrak{osp}(1|2)$ no example does exist except for one that annihilates the vacuum vector.
Then the problem is whether there is an example for another Lie superalgebra of basic classical type.
In seeking an example of an annihilating operator, one has to notice that
the form of an annihilator Eq.(\ref{eq:annihilator}) has room of generalization.
Indeed one may make the parameter $\tau$ depend on the summation index,
which is realized by using Brownian motions of different covariance along internal degrees of freedom
allowing inhomogeneity in an internal space.
One can also use a term $X(-2)$ for $X\in\mfrak{g}$ in an annihilator.
Such a deformation is thought to be inevitable when we twist the Virasoro field by a derivative of a current field in the Sugawara construction
as is checked on the vacuum representation.
Note that current field is not primary with respect to a twisted Virasoro field,
which forces us to change the semi-direct product structure of the target group to formulate SLE.
In conclusion, we suggest that the most general form of an annihilator becomes
\begin{equation}
	-2L_{-2}+\frac{\kappa}{2}L_{-1}^{2}+c X(-2)+\frac{1}{2}\sum_{a=1}^{\dim\mfrak{g}}(-1)^{p(X_{a})}\tau_{a}X_{a}(-1)X^{a}(-1)
\end{equation}
where $\kappa, \tau_{a}>0$, $c\in\mbb{C}$ and $X\in\mfrak{g}$.
It is not obvious whether an annihilating operator of this form exists except for one on the vacuum representation
even for the case of $\mfrak{g}=\mfrak{osp}(1|2)$.

We may also look for an annihilating operator of a highest weight vector with degree higher than 2.
It is known that there is a degree 4 operator that annihilates the vacuum vector for the Yang-Lee singularity
and an SLE-type growth process corresponding to this annihilating operator was considered\cite{LesageRasmussen2004}.
An analogous annihilating operator of degree 4 was also found for affine $\mfrak{sl}_{2}$ and $\mfrak{sl}_{3}$,
to which certain SLE-type growth processes with internal symmetry correspond\cite{SK2017}.
It may be interesting to find an annihilating operator of a highest weight vector with higher degree for an affine Lie superalgebra
in that it will broaden applicability of the formulation presented in this paper.
Note that assuming existence of an annihilating operator that is realized as at most quadratic in generators of an algebra,
it is possible to write down SDEs associated to the operator.
We also remark that in the case of an affine Lie algebra\cite{SK2017},
an annihilating operator of degree 4 was found with the help of the Frenkel-Kac construction\cite{FrenkelKac1980},
of which the equivalent to an affine Lie superalgebra is not clear at least to us.

Other possibly interesting Lie superalgebras than basic classical ones we considered in this paper
include the affinization of supercommutative Lie superalgebras.
For such cases, the annihilator conditions will be investigated explicitly.
Construction of random process along internal degrees of freedom and computation of local martingales will also be carried out
for general rank of the underlying Lie superalgeras.
Such examples are super-analogues of SLE corresponding to Heisenberg algebras considered in our previous work\cite{SK2018a}.

In a literature \cite{Sakai2013}, the author formulated multiple SLEs with internal symmetry corresponding to WZW theories.
It will be interesting to extend the result of the present paper to multiple SLEs.
Note that a formulation of multiple SLEs\cite{BauerBernardKytola2005,Sakai2013} relies on correlation functions,
and local martingales are obtained as correlation functions normalized by partition functions,
which are also correlation functions among boundary fields.
In the formulation of the present paper, we obtain local martingales after integrating Grassmann variables.
We will need the correct analogy of this prescription in the formulation via correlation functions
to extend our result to multiple SLEs.

\begin{figure}[h]
\centering
\includegraphics[width=5cm]{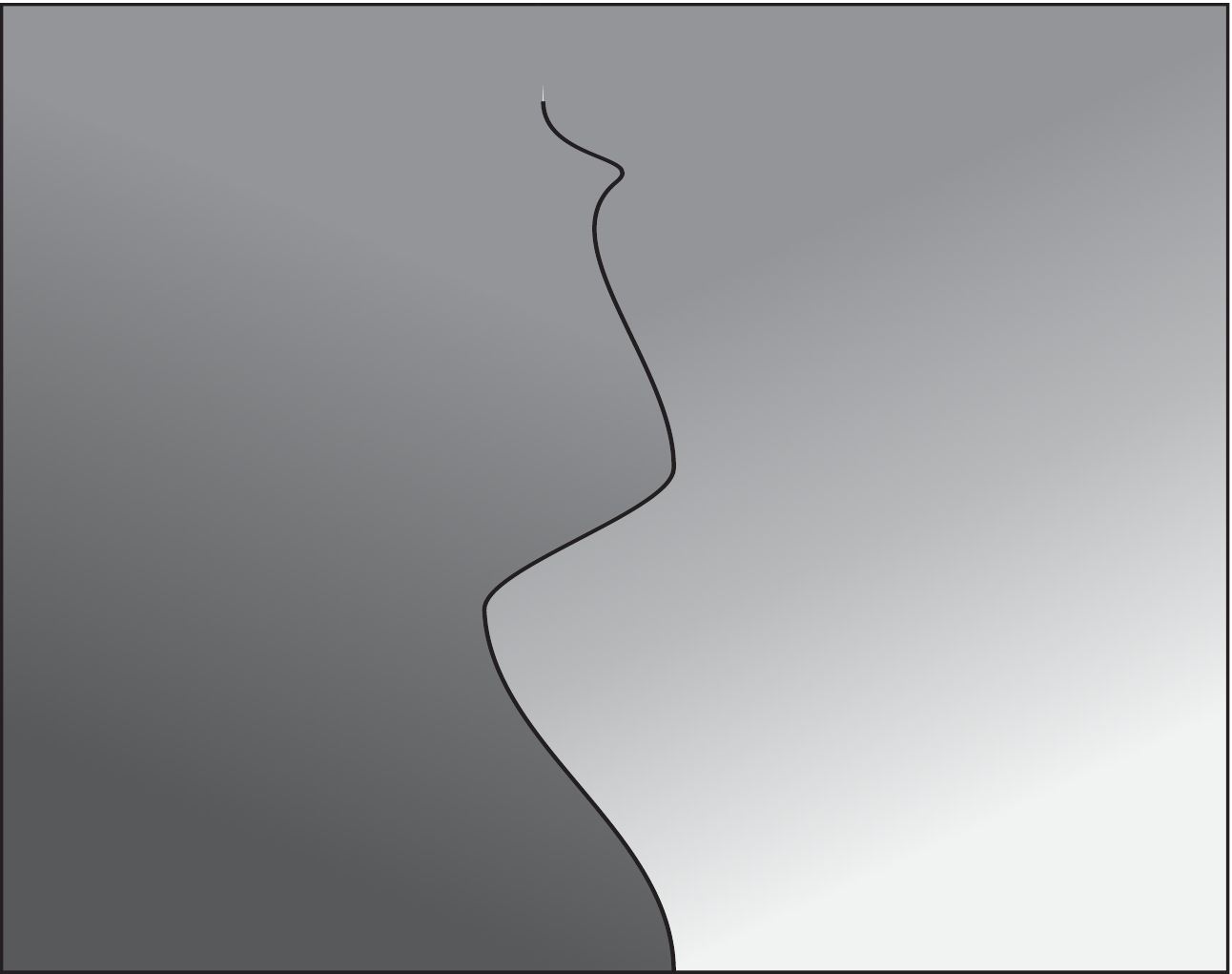}
\caption{SLE with internal symmetry}
\label{fig:SLE_internal}
\end{figure}

One of the most important applications of SLE/CFT correspondence is a generalization of Cardy's formula.
Indeed, due to a local martingale that is computed from a representation of the Virasoro algebra,
Cardy's formula was rederived \cite{BauerBernard2003a,BauerBernard2006}.
Thus a generalization of SLE that corresponds to some CFT has potential to propose a generalization of Cardy's formula.
To seek a statistical model that is related to our generalized SLE,
together with a generalization of Cardy's formula, will be a significant direction of future research.
Related to this, we shall comment that a variant of SLE with internal symmetry can be considered as
a complexified Bessel process with internal degrees of freedom 
when we evaluate the coordinate $z$ and the affine parameter $\zeta$ at any point in the upper half plane.
This observation allows one to identify each realization of SLE with internal symmetry
with an evolution of a curve in the upper half plane and coloring of the plane as illustrated in Fig.\ref{fig:SLE_internal}.
In this interpretation, the relevant Lie supergroup acts on the coloring reflecting the Lie superalgebra symmetry of the model.
This is expected to serve as a starting point of relating SLE with internal symmetry to physical systems.

\addcontentsline{toc}{chapter}{Bibliography}
\bibliographystyle{alpha}
\bibliography{sle_cft}

\newcommand{\etalchar}[1]{$^{#1}$}
\begin{thebibliography}{CDCH{\etalchar{+}}14}

\bibitem[ABI11]{AlekseevBytskoIzyurov2011}
A.~Alekseev, A.~Bytsko, and K.~Izyurov.
\newblock On {SLE} martingales in boundary {WZW} models.
\newblock {\em Lett. Math. Phys.}, 97:243--261, 2011.

\bibitem[BB02]{BauerBernard2002}
M.~Bauer and D.~Bernard.
\newblock {SLE}$_{\kappa}$ growth processes and conformal field theories.
\newblock {\em Phys. Lett. B}, 543:135--138, 2002.

\bibitem[BB03a]{BauerBernard2003a}
M.~Bauer and D.~Bernard.
\newblock Conformal field theories of stochastic {Loewner} evolutions.
\newblock {\em Commun. Math. Phys.}, 239:493--521, 2003.

\bibitem[BB03b]{BauerBernard2003b}
M.~Bauer and D.~Bernard.
\newblock {SLE} martingales and the {Virasoro} algebra.
\newblock {\em Phys. Lett. B}, 557:309--316, 2003.

\bibitem[BB06]{BauerBernard2006}
M.~Bauer and D.~Bernard.
\newblock {2D} growth processes: {SLE} and {Loewner} chains.
\newblock {\em Phys. Rep.}, 432:115--221, 2006.

\bibitem[BBK05]{BauerBernardKytola2005}
M.~Bauer, D.~Bernard, and K.~Kyt{\"o}l{\"a}.
\newblock Multiple {Schramm-Loewner} evolutions and statistical mechanics
  martingales.
\newblock {\em J. Stat. Phys.}, 120:1125--1163, 2005.

\bibitem[Ber95]{Bernard1995}
D.~Bernard.
\newblock (perturbed) conformal field theory applied to {2D} disordered
  systems: an introduction, 1995.
\newblock arXiv:hep-th/9509137.

\bibitem[BGLW05]{BettelheimGruzbergLudwigWiegmann2005}
E.~Bettelheim, I.~A. Gruzberg, A.~W.~W. Ludwig, and P.~Wiegmann.
\newblock Stochastic {Loewner} evolution for conformal field theories with
  {Lie} group symmetries.
\newblock {\em Phys. Rev. Lett.}, 95:251601, 2005.

\bibitem[BPZ84]{BelavinPolyakovZamolodchikov1984}
A.~A. Belavin, A.~M Polyakov, and A.~B. Zamolodchikov.
\newblock Infinite conformal symmetry in two-dimensional quantum field theory.
\newblock {\em Nucl. Phys. B}, 241:333--380, 1984.

\bibitem[BT81]{BerezinTolstoy1981}
F.~A. Berezin and V.~N. Tolstoy.
\newblock The group with {Grassmann} structure {$UOSP(1,2)$}.
\newblock {\em Commun. Math. Phys.}, 78:409--428, 1981.

\bibitem[Car92]{Cardy1992}
J.~L. Cardy.
\newblock Critical percolation in finite geometries.
\newblock {\em J. Phys. A: Math. Gen.}, 25:L201--L206, 1992.

\bibitem[CDCH{\etalchar{+}}14]{ChelkakDuminil-CopinHonglerKemppainenSmirnov2014}
D.~Chelkak, H.~Duminil-Copin, C.~Hongler, A.~Kemppainen, and S.~Smirnov.
\newblock Convergence of {Ising} interfaces to {Schramm's SLE} curves.
\newblock {\em Comptes Rendus Mathematique}, 352:157--161, 2014.

\bibitem[CW13]{ChengWang2013}
S.~Cheng and W.~Wang.
\newblock {\em Dualities and Representations of Lie Superalgebras}, volume 114
  of {\em Graduate Studies in Mathematics}.
\newblock American Mathematical Society, 2013.

\bibitem[Dub15a]{Dubedat2015b}
J.~Dub{\'e}dat.
\newblock {SLE} and {Virasoro} representations: Fusion.
\newblock {\em Commun. Math. Phys.}, 336:761--809, 2015.

\bibitem[Dub15b]{Dubedat2015a}
J.~Dub{\'e}dat.
\newblock {SLE} and {Virasoro} representations: Localization.
\newblock {\em Commun. Math. Phys.}, 336:695--760, 2015.

\bibitem[Efe83]{Efetov1983}
K.B. Efetov.
\newblock Supersymmetry and theory of disordered metals.
\newblock {\em Advances in Physics}, 32:53--127, 1983.

\bibitem[EFS05]{EsslerFrahmSaleur2005}
F.~H.~L. Essler, H.~Frahm, and H.~Saleur.
\newblock Continuum limit of the integrable $sl(2/1)\ 3-\bar{3}$ superspin
  chain.
\newblock {\em Nucl. Phys. B}, 712:513--572, 2005.

\bibitem[FBZ04]{FrenkelBen-Zvi2004}
E.~Frenkel and D.~Ben-Zvi.
\newblock {\em Vertex Algebras and Algebraic Curves}, volume~88 of {\em
  Mathematical Surveys and Monographs}.
\newblock American Methematical Society, 2nd edition, 2004.

\bibitem[FK80]{FrenkelKac1980}
I.~B. Frenkel and V.~G. Kac.
\newblock Basic representations of affine {Lie} algebras and dual resonance
  models.
\newblock {\em Invent. Math.}, 62:23--66, 1980.

\bibitem[FK04]{FriedrichKalkkinen2004}
R.~Friedrich and J.~Kalkkinen.
\newblock On conformal field theory and stochastic {Loewner} evolution.
\newblock {\em Nucl. Phys. B}, 687:279--302, 2004.

\bibitem[Fri04]{Friedrich2004}
R.~Friedrich.
\newblock On connections of conformal field theory and stochastic {Loewner}
  evolution, 2004.
\newblock arXiv:math-ph/0410029.

\bibitem[FSS00]{FrappatSciarrinoSorba2000}
L.~Frappat, A.~Sciarrino, and P.~Sorba.
\newblock {\em Dictionary on Lie Algebras and Superalgebras}.
\newblock Academic Press, 2000.
\newblock arXiv:hep-th/9607161.

\bibitem[FW03]{FriedrichWerner2003}
R.~Friedrich and W.~Werner.
\newblock Conformal restriction, highest-weight representations and {SLE}.
\newblock {\em Commun. Math. Phys.}, 243:105--122, 2003.

\bibitem[Kon03]{Kontsevich2003}
M.~Kontsevich.
\newblock {CFT, SLE} and phase boundaries, 2003.
\newblock Oberwolfach Arbeitstagung.

\bibitem[Kos17]{SK2017}
S.~Koshida.
\newblock {SLE}-type growth processes corresponding to {Wess-Zumino-Witten}
  theories, 2017.
\newblock arXiv:1710.03835.

\bibitem[Kos18]{SK2018a}
S.~Koshida.
\newblock Local martingales associated with {SLE} with internal symmetry, 2018.
\newblock arXiv:1803.06808.

\bibitem[Kyt07]{Kytola2007}
K.~Kyt{\"o}l{\"a}.
\newblock Vorasoro module structure of local martingales of {SLE} variants.
\newblock {\em Rev. Math. Phys.}, 5:455--509, 2007.

\bibitem[KZ84]{KnizhnikZamolodchikov1984}
V.~G. Knizhnik and A.~B. Zamolodchikov.
\newblock Current algebra and {Wess-Zumino} model in two dimensions.
\newblock {\em Nuclear Physics B}, 247:83--103, 1984.

\bibitem[Law04]{Lawler2004}
G.~F. Lawler.
\newblock An introduction to the stochastic {Loewner} evolution.
\newblock In {\em Random Walks and Geometry}. De Gruyter, 2004.

\bibitem[LR04]{LesageRasmussen2004}
F.~Lesage and J.~Rasmussen.
\newblock {SLE}-type growth processes and the {Yang-Lee} singularity.
\newblock {\em J. Math. Phys.}, 45:3040--3048, 2004.

\bibitem[MARR04]{Moghimi-AraghiRajabpourRouhani2004}
A.~Moghimi-Araghi, M.~A. Rajabpour, and S.~Rouhani.
\newblock Logarithmic conformal null vectors and {SLE}.
\newblock {\em Phys. Lett. B}, 600:298--301, 2004.

\bibitem[Naz12]{Nazarov2012}
A.~Nazarov.
\newblock {Schramm-Loewner} evolution martingales in coset conformal field
  theory.
\newblock {\em JETP Letters}, 96:90--93, 2012.

\bibitem[NR05]{NagiRasmussen2005}
J.~Nagi and J.~Rasmussen.
\newblock On stochastic evolutions and superconformal field theory.
\newblock {\em Nucl. Phys. B}, 704:475--489, 2005.

\bibitem[Ras04a]{Rasmussen2004b}
J.~Rasmussen.
\newblock Note on stochastic {L}{\"o}wner evolutions and logarithmic conformal
  field theory.
\newblock {\em J. Stat. Mech.}, page P09007, 2004.

\bibitem[Ras04b]{Rasmussen2004a}
J.~Rasmussen.
\newblock Stochastic evolutions in superspace and superconformal field theory.
\newblock {\em Lett. Math. Phys.}, 68:41--52, 2004.

\bibitem[Ras07]{Rasmussen2007}
J.~Rasmussen.
\newblock On {$SU(2)$ Wess-Zumino-Witten} models and stochastic evolutions.
\newblock {\em Afr. J. Math. Phys.}, 4:1--9, 2007.

\bibitem[RS05]{RohdeSchramm2005}
S.~Rohde and O.~Schramm.
\newblock Basic properties of {SLE}.
\newblock {\em Ann. Math.}, 161:883--924, 2005.

\bibitem[Sak13]{Sakai2013}
K.~Sakai.
\newblock Multiple {Schramm-Loewner} evolutions for conformal field theories
  with {Lie} algebra symmetries.
\newblock {\em Nucl. Phys. B}, 867:429--447, 2013.

\bibitem[Sch00]{Schramm2000}
O.~Schramm.
\newblock Scaling limits of loop-erased random walks and uniform spanning
  trees.
\newblock {\em Israel J. Math.}, 118:221--288, 2000.

\bibitem[Smi01]{Smirnov2001}
S.~Smirnov.
\newblock Critical percolation in the plane: conformal invariance, {Cardy's}
  formula, scaling limits.
\newblock {\em C. R. Acad. Sci. Paris}, 333:239--244, 2001.

\bibitem[SS07]{SaleurSchomerus2007}
H.~Saleur and V.~Schomerus.
\newblock On the {$SU(2|1)$} {WZNW} model and its statistical mechanics
  applications.
\newblock {\em Nucl. Phys. B}, 775:312--340, 2007.

\bibitem[Var04]{Varadarajan2004}
V.~S. Varadarajan.
\newblock {\em Supersymmetry for Mathematicians: An Introduction}, volume~11 of
  {\em Courant Lecture Notes in Mathematics}.
\newblock American Mathematical Society, Providence, RI, 2004.

\bibitem[Wak01]{Wakimoto2001}
M.~Wakimoto.
\newblock {\em Lectures on Infinite-Dimensional Lie Algebra}.
\newblock World Scientific Publishing Co. Pte. Ltd., 2001.

\bibitem[Wer03]{Werner2003}
W.~Werner.
\newblock Random planar curves and {Schramm-Loewner} evolutions, 2003.
\newblock arXiv:math/0303354.

\bibitem[Wit84]{Witten1984}
E.~Witten.
\newblock Non-abelian bosonization in two dimensions.
\newblock {\em Commun. Math. Phys.}, 92:455--472, 1984.

\bibitem[WZ71]{WessZumino1971}
J.~Wess and B.~Zumino.
\newblock Consequences of anomalous {Ward} identity.
\newblock {\em Phys. Lett. B}, 37:95--97, 1971.

\end{thebibliography}
\end{document}